\documentclass{llncs}

\title{Extending Prolog with \\ Incomplete Fuzzy Information}

\author{  Susana Munoz-Hernandez \inst{1}  \and 
          Claudio Vaucheret\inst{2} }

 \institute{ Departamento de Lenguajes, Sistemas de la Informaci\'{o}n
                e Ingenier\'{\i}a del Software \\
                Facultad de
                Inform\'{a}tica - Universidad Polit\'{e}cnica de
                Madrid \\ Campus de Montegancedo 28660 Madrid, Spain \\
		\email{susana@fi.upm.es} 
                \and
                Departamento de Ciencias de la Computaci\'{o}n \\
                Facultad de Econom\'{\i}a y Administraci\'{o}n \\ 
                Universidad Nacional del
                Comahue \\  Universidad Polit\'{e}cnica de
                Madrid \\ Buenos Aires 1400 (8300) Neuquen, Argentina
                \\ 
                \email{vaucheret@ibap.com.ar }}

\usepackage{amssymb}

\usepackage{graphicx} 

\usepackage{pst-node}

\newcommand{\AmS}{{\protect\the\textfont2
  A\kern-.1667em\lower.5ex\hbox{M}\kern-.125emS}}

\newtheorem{defn}{Definition}[section]
\newtheorem{thm}{Theorem}[section]
\newtheorem{lem}{Lemma}[section]
\newenvironment{pf}{\noindent \emph{Proof.} }{}

\newtheorem{rem}{Remark}[section]

\begin{document}

\maketitle

\begin{abstract}
  Incomplete information is a problem in many aspects of actual environments.
  Furthermore, in many sceneries the knowledge is not represented in a crisp
  way. It is common to find fuzzy concepts or problems with some level of
  uncertainty. There are not many practical systems which handle
  fuzziness and uncertainty and the few examples that we can find are used by
  a minority. To extend a popular system (which many programmers are
  using) with the ability of combining crisp and fuzzy knowledge
  representations seems to be an interesting issue.

  Our first work (Fuzzy Prolog
  ) was a language that models
  $\mathcal{B}([0,1])$-valued Fuzzy Logic. In the Borel algebra,
  $\mathcal{B}([0,1])$, truth value is represented using unions of intervals
  of real numbers. This work was more general
  in truth value representation and propagation than previous works.


  An interpreter for this language using Constraint Logic Programming over
  Real numbers (CLP(${\cal R}$)) was implemented and is available in the Ciao
  system 
  .
  
  Now, we enhance our former approach by using default knowledge to represent
  incomplete information in Logic Programming. We also provide the
  implementation of this new framework.  This new release of Fuzzy Prolog
  handles incomplete information, it has a complete semantics (the previous one
  was incomplete as Prolog) and moreover it is able to combine crisp and fuzzy
  logic in Prolog programs. Therefore, new Fuzzy Prolog is more expressive to
  represent real world.

  Fuzzy Prolog inherited from Prolog its incompleteness. The incorporation of
  default reasoning to Fuzzy Prolog removes this problem and requires a richer
  semantics which we discuss.


\end{abstract}



\newcommand{\proyecto}{TIC2003-01036}

\renewcommand{\thefootnote}{\ }
\footnotetext{ 
 
\noindent
This research was partly supported by the Spanish MCYT project \proyecto.
}

\renewcommand{\thefootnote}{\arabic{footnote}}

\section{Introduction}
\label{sec:intr}

World information is not represented in a crisp way. Its
representation is imperfect, fuzzy, etc., so that the management of
uncertainty is very important in knowledge representation. There are
multiple frameworks for incorporating uncertainty in logic
programming:
\begin{itemize}
     \item fuzzy set theory \cite{Cao00,Shapiro83,Emden86},

     \item probability theory
     \cite{Fuhr00,LakshmananShiri94,Lukasiewicz01,Ng91,Ng93},

     \item multi-valued logic
     \cite{Fitting91,Kifer88,Kifer92,Lakshmanan94,Lakshmanan01,Subrahmanian87},

     \item possibilistic logic \cite{Dubois91,Wagner97,Wagner98}
\end{itemize}

In \cite{Lakshmanan01} a general framework was proposed that generalizes many
of the previous approaches. At the same time an analogous theoretical
framework was provided and a prototype for Prolog was
implemented~\cite{CUW01}.  Basically, a rule is of the form $A \leftarrow
B_1,\ldots,B_n$, where the assignment $I$ of certainties is taken from a
certainty lattice, to the $B_i$s. The certainty of $A$ is computed by taking
the set of the certainties $I(B_i)$ and then they are propagated using the
function $F$ that is an aggregation operator. This is a very flexible approach
and in \cite{Vaucheret_LPAR02,Susana_FSS04} practical examples in a Prolog
framework are presented.

In this work we extend the approach of \cite{Susana_FSS04} with
arbitrary assignments of default certainty values (non-uniform default
assumptions). The usual semantics of logic programs can be obtained
through a unique computation method, but using different
assumptions in a uniform way to assign the same default truth-value to
all the atoms. The most well known assumptions are:
\begin{itemize}

    \item the \emph{Closed World Assumption} (CWA), which asserts that
any atom whose truth-value cannot be inferred from the facts and
clauses of the program is supposed to be false (i.e. certainty
$0$). It is used in stable models \cite{Gelfond,Gelfond90} and
well-founded semantics \cite{Loyer02,Lukasiewicz01,Ng91},

     \item the \emph{Open World Assumption} (OWA), which asserts that any
atom whose truth-value cannot be inferred from the facts and clauses
of the program is supposed to be undefined or unknown (i.e. certainty
in $[0,1]$). It is used in \cite{Lakshmanan01}.

\end{itemize}

There are also some approaches \cite{Wagner97,Wagner98} where both
assumptions can be combined and some atoms can be interpreted assuming
CWA while others follows OWA. Anyway, what seems really interesting is
not only to combine both assumptions but to generalize the use of a
default value. The aim is working with incomplete information with
more guarantees.


The rest of the paper is organized as follows. Section
\ref{sec:fuzzyprolog} introduces the Fuzzy Prolog language. A complete
description of the new semantics of Fuzzy Prolog is provided in Section
\ref{sec:semantics}. Section \ref{sec:clpr} completes the details
about the improved implementation using CLP(${\cal R}$) with the extension to
handle default knowledge. 
Some illustrating examples are provided in section \ref{sec:comb}.
Finally, we conclude and discuss some future work (Section \ref{sec:concl}).

\section{Fuzzy Prolog}
\label{sec:fuzzyprolog}

In this section we are going to summarize the main characteristics of the Fuzzy
Prolog that we proposed in \cite{Susana_FSS04} and that is the basis of the
work presented here. Fuzzy Prolog is more general than
previous approaches to introduce fuzziness in Prolog in some respects:
\begin{enumerate}
  
\item A truth value will be a finite union of closed sub-intervals on
  $[0,1]$. This is represented by Borel algebra, $\mathcal{B}([0,1])$, while
  the algebra $\mathcal{E}([0,1])$ only considers intervals. An interval is a
  special case of the union of one element, and a unique truth value is a
  particular case of having an interval with only one element.

\item A truth value will be propagated through the rules by means of
  an \emph{aggregation operator}. The definition of \emph{aggregation
  operator} is general in the sense that it subsumes conjunctive
  operators (triangular norms \cite{Norms00} like min, prod, etc.),
  disjunctive operators \cite{Tri_Cub_Cas}(triangular co-norms, like
  max, sum, etc.), average operators (like arithmetic average,
  quasi-linear average, etc) and hybrid operators (combinations of the
  above operators \cite{Prad_Tri_Cal}).

\item The declarative and procedural semantics for
Fuzzy Logic programs are given and their equivalence is proven.

\item An implementation of the proposed language is presented. A \emph{fuzzy program} is a finite set of 
\begin{itemize}
    \item \emph{fuzzy facts} ($A \gets v$, where $A$ is an atom and
        $v$, a truth value, is an element in $\mathcal{B}([0,1])$
        expressed as constraints over the domain $[0,1]$), and

     \item \emph{fuzzy clauses} ($A \gets_F B_1,\ldots,B_n$, where
$A,B_1,\ldots,B_n$ are atoms, and $F$ is an interval-aggregation
operator, which induces a union-aggregation, as by Definition
\ref{def:Faggr}, $\mathcal{F}$ of truth values in $\mathcal{B}([0,1])$
represented as constraints over the domain $[0,1]$).

\end{itemize}

     We obtain information from the program through \emph{fuzzy
queries or fuzzy goals} ($ v \gets A ~?$ where $A$
is an atom, and $v$ is a variable, possibly instantiated, that
represents a truth value in $\mathcal{B}([0,1])$).
\end{enumerate}

Programs are defined as usual but handling
truth values in $\mathcal{B}([0,1])$ (the Borel algebra over the real
interval $[0,1]$ that deals with unions of intervals) represented as
constraints. We refer, for example, to expressions as: $(v \geq 0.5 ~
\wedge ~ v \leq 0.7) ~\vee~ (v \geq 0.8 ~ \wedge ~ v \leq 0.9)$ to
represent a truth value in $[0.5, 0.7] ~ \bigcup ~ [0.8, 0.9]$.

A lot of everyday situations can only be represented by this general
representation of truth value. There are some examples in
\cite{Susana_FSS04}.


The truth value of a goal will depend on the truth value of the subgoals which
  are in the body of the clauses of its definition. Fuzzy Prolog
  \cite{Susana_FSS04} uses \emph{aggregation operators} \cite{Tri_Pra_Cub} in
  order to propagate the truth value by means of the fuzzy rules.  Fuzzy sets
  \emph{aggregation} is done using the application of a numeric operator of
  the form $f: [0,1]^n \to [0,1]$. An \emph{aggregation operator} must verify
  $f(0,\ldots,0) = 0$ and $f(1,\ldots,1) = 1$, and in addition it should be
  monotonic and continuous. If we deal with the definition of fuzzy sets as
  intervals it is necessary to generalize from \emph{aggregation operators} of
  numbers to \emph{aggregation operators} of intervals. Following the theorem
  proven by Nguyen and Walker in \cite{nguyen1} to extend T-norms and
  T-conorms to intervals, we propose the following definitions.

\begin{defn}[interval-aggregation]
Given an aggregation $f: [0,1]^n \to [0,1]$, an
interval-aggregation $F: \mathcal{E}([0,1])^n \to
\mathcal{E}([0,1])$ is defined as follows:
\[ F([x^l_1,x^u_1],...,[x^l_n,x^u_n]) =
[f(x^l_1,...,x^l_n),f(x^u_1,...,x^u_n)].\]
\end{defn}

Actually, we work with union of intervals and propose the
definition:

\begin{defn}[union-aggregation]
\label{def:Faggr} Given an interval-aggregation \\
$F:\mathcal{E}([0,1])^n \to \mathcal{E}([0,1])$ defined
over intervals, a union-aggregation \\ $\mathcal{F}:
\mathcal{B}([0,1])^n \to \mathcal{B}([0,1])$ is defined
over union of intervals as follows: $$\mathcal{F}(B_1,\ldots,B_n)
= \cup \{ F(\mathcal{E}_1,...,\mathcal{E}_n) ~|~ \mathcal{E}_i \in
B_i \}.$$
\end{defn}

In the presentation of the theory of possibility \cite{Zad1},
Zadeh considers that fuzzy sets act as an elastic constraint on
the values of a variable and fuzzy inference as constraint
propagation.

In \cite{Susana_FSS04} (and furthermore in the extension that we presented in
this paper), truth values
and the result of aggregations are represented by constraints. A
constraint is a \emph{$\Sigma$-formula} where $\Sigma$ is a signature
that contains the real numbers, the binary function symbols $+$ and
$*$, and the binary predicate symbols $=$, $<$ and $\leq$. If the
constraint $c$ has solution in the domain of real numbers in the
interval $[0,1]$ then $c$ is \emph{consistent}, and is denoted as
$\mathit{solvable}(c)$.


\section{Semantics}
\label{sec:semantics}

This section contains a reformulation of the semantics of Fuzzy Prolog. This
new semantics is complete thanks to the inclusion of default value.
\subsection{Least Model Semantics}
\label{sec:sem}

The \emph{Herbrand universe} $U$ is the set of all ground
\emph{terms}, which can be made up with the constants and function
symbols of a program, and the \emph{Herbrand base} $B$ is the set
of all ground atoms which can be formed by using the predicate
symbols of the program with ground \emph{terms} (of the
\emph{Herbrand universe}) as arguments.


\begin{defn}[default value]
  We assume there is a function \emph{default} which implement the Default
  Knowledge Assumptions. It assigns an element of $\mathcal{B}([0,1])$
  to each element of the Herbrand Base. If the Closed World Assumption
  is used, then $\mathit{default}(A) = [0,0]$ for all $A$ in Herbrand Base. If
  Open World Assumption is used instead, $\mathit{default}(A) = [0,1]$ for all
  $A$ in Herbrand Base.
\end{defn}

\begin{defn}[interpretation]
An \emph{interpretation} $I= \langle B_I,V_I \rangle $ consists of the
following:
 \begin{enumerate}
        \item a subset $B_I$ of the \emph{Herbrand Base},
        \item a mapping $V_I$, to assign 
          \begin{enumerate}
          \item a truth value, in
        $\mathcal{B}([0,1])$, to each element of $B_I$, or
      \item $\mathit{default}(A)$, if $A$ does not belong to $B_I$.
          \end{enumerate}
 \end{enumerate}
\end{defn}

\begin{defn}[interval inclusion $\subseteq_{II}$]
  Given two intervals $I_1 = [a,b]$, $I_2 = [c,d]$ in
  $\mathcal{E}([0,1])$, $I_1 \subseteq_{II} I_2$ if and only if $c
  \leq a$ and $b \leq d$.
\end{defn}

\begin{defn}[Borel inclusion $\subseteq_{BI}$]
 Given two unions of intervals $U = I_1 \cup \dots \cup I_N$, $U' = I_1' \cup
  \dots \cup I_M'$ in $\mathcal{B}([0,1])$, $U \subseteq_{BI} U'$ if and only
  if $\forall I_i \in U$, $i \in {1..N}$, $I_i$ can be partitioned in to
  intervals $J_{i1}, ..., J_{iL}$, i.e.  
  $J_{i1} \cup ... \cup J_{iL} = I_i, ~J_{i1} \cap ... \cap J_{iL}$ is the set
  of the border elements of the intervals except the lower limit of $J_{i1}$
  and the upper limit of $J_{iL}$) and for all $k \in {1..L},~ \exists J_{jk}'
  \in U' ~.~ J_{ik} \subseteq_{II} J_{jk}'$ where $jk \in {1..M}$.
\end{defn}

The Borel algebra $\mathcal{B}([0,1])$ is a complete lattice under
$\subseteq_{BI}$ (Borel inclusion), and the Herbrand
base is a complete lattice under $\subseteq$ (set
inclusion) and so the set of all \emph{interpretations} forms a
complete lattice under the relation $\sqsubseteq$ defined as
follows.

\noindent
Notice that we have redefined interpretation and Borel inclusion with respect
to the definitions in \cite{Susana_FSS04}. We will also redefine the
operational semantics and therefore the internal implementation of the Fuzzy
Prolog library.  Sections below are completely new too. For uniformity reasons
we have kept the same syntax that was used in \cite{Susana_FSS04} in fuzzy
programs.

\begin{defn}[interpretation inclusion $\sqsubseteq$]
  Let $I = \langle B_I,V_I \rangle$ and \\ $I' = \langle B_{I'},V_{I'} \rangle$
  be interpretations. $I \sqsubseteq I'$ if and only if $B_I \subseteq B_{I'}$
  and for all $B \in B_I$, $V_I(B) \subseteq_{BI} V_{I'}(B)$.
\end{defn}

\begin{defn}[valuation]
  A \emph{valuation} $\sigma$ of an atom $A$ is an assignment of
  elements of $U$ to variables of $A$. So $\sigma(A) \in B$ is a
  ground atom.
\end{defn} 
In the Herbrand context, a valuation is the same as a substitution.

\begin{defn}[model]
  Given an \emph{interpretation} $I = \langle B_I,V_I \rangle$,
  \begin{itemize}
    \item $I$ is a \emph{model} for a \emph{fuzzy fact} $A \gets v$,
    if for all valuations $\sigma$, $\sigma(A) \in B_I \mbox{ and }$ $ v
    \subseteq_{BI} V_I(\sigma(A))$.

  \item I is a \emph{model} for a clause $A \gets_F B_1,\ldots,B_n$
    when the following holds: for all valuations $\sigma$, $\sigma(A)
    \in B_I$ and $v \subseteq_{BI} V_I(\sigma(A))$, where $v =
    \mathcal{F}(V_I(\sigma(B_1)),\ldots,$ $V_I(\sigma(B_n)))$ and
    $\mathcal{F}$ is the union aggregation obtained from $F$.

    \item $I$ is a \emph{model} of a \emph{fuzzy program}, if it is a
      \emph{model} for the facts and clauses of the program.
  \end{itemize}
\end{defn}

Every program has a least model which is usually regarded as the intended
interpretation of the program since it is the most conservative model.  Let
$\cap$ (that appears in the following theorem) be the meet operator on the
lattice of interpretations $(I,\sqsubseteq)$. We can prove the following
result.

  \begin{thm}[model intersection property]
    Let $I_1 = \langle B_{I_1},V_{I_1} \rangle$, ~$I_2 = \langle
    B_{I_1},V_{I_1} \rangle$ be models of a fuzzy program $P$. Then
    $I_1 \cap I_2$ is a model of $P$.
  \end{thm}

  \begin{pf}
    Let $M = \langle B_M,V_M \rangle =I_1 \cap I_2$.
    Since $I_1$ and $I_2$ are models of $P$, they are models for each
    fact and clause of $P$. Then for all valuations $\sigma$ we have
    \begin{itemize}
    \item for all facts $A \gets v$ in $P$,
      \begin{itemize}
      \item $\sigma(A) \subseteq B_{I_1}$ and $\sigma(A) \in
        B_{I_2}$, and so $\sigma(A) \in B_{I_1} \cap B_{I_2}= B_M$,
      \item $v \subseteq_{BI} V_{I_1}(\sigma(A))$ and $v
      \subseteq_{BI} V_{I_2}(\sigma(A))$, and so hence \\ $v \subseteq_{BI}
      V_{I_1}(\sigma(A)) \cap V_{I_2}(\sigma(A))= V_M(\sigma(A))$
      \end{itemize}
      therefore $M$ is a model for $A \gets v$
    \item  and for all clauses $A \gets_F B_1,\ldots,B_n$ in $P$
      \begin{itemize}
      \item since $\sigma(A) \in B_{I_1}$ and $\sigma(A) \in B_{I_2}$, hence $\sigma(A)
      \in B_{I_1} \cap B_{I_2}= B_M$.
    \item if $v =
      \mathcal{F}(V_M(\sigma(B_1)),\ldots,V_M(\sigma(B_n)))$, since
      $F$ is monotonic, $v
      \subseteq_{BI} V_{I_1}(\sigma(A))$ and $v \subseteq_{BI}
      V_{I_2}(\sigma(A))$, hence $v \subseteq_{BI} V_{I_1}(\sigma(A))
      \cap V_{I_2}(\sigma(A))= V_M(\sigma(A))$
      \end{itemize}
      therefore $M$ is a model for $A \gets_F B_1,\ldots,B_n$
    \end{itemize}
    and $M$ is model of $P$.
  \end{pf}

\begin{rem}[Least model semantic]
If we let $\mathbf{M}$ be the set of all models of a program $P$,
the intersection of all of these models, $\bigcap \mathbf{M}$, is a
model and it is the least model of $P$. We denote the least model
of a program $P$ by $lm(P)$.
\end{rem}

\subsection{Fixed-Point Semantics}
\label{sec:fixp}

The fixed-point semantics we present is based on a one-step
consequence operator $T_P$. The least fixed-point $\mathit{lfp}(T_P)=I$
(i.e. $T_P(I) = I$) is the declarative meaning of the program $P$,
so is equal to $lm(P)$. We include it here for clarity reasons although it is
the same that in \cite{Susana_FSS04}.

Let $P$ be a fuzzy program and $B_P$ the Herbrand base of $P$;
then the \emph{mapping} $T_P$ over \emph{interpretations} is
defined as follows:

Let $I = \langle B_I,V_I \rangle$ be a fuzzy \emph{interpretation}, then
$T_P(I) = I'$, $I' = \langle B_{I'},V_{I'} \rangle$, $B_{I'} = \{ A \in B_P
~|~ Cond \}, V_{I'}(A) = \bigcup \{ v \in \mathcal{B}([0,1]) ~|~ Cond \} $

where $$
\begin{array}{rl}

Cond = & ( A \gets v \mbox{ is a ground instance} 
   \mbox{ of a fact in } P \mbox{ and } \mathit{solvable}(v) ) \\
  & \mbox{ or } \\
  & ( A \gets_F A_1,\ldots,A_n \mbox{ is a ground}
   \mbox{ instance of a clause in } P,  \\
  & \mbox{ and } \mathit{solvable}(v),
   v = \mathcal{F}(V_I(A_1),\ldots,V_I(A_n)) ). \\
\end{array} $$

\noindent Note that since
$I'$ must be an interpretation, $V_{I'}(A)= \mathit{default}(A)$ for all $A \notin
B_{I'}$.

The set of interpretations forms a complete lattice, so that
$T_P$ it is continuous.

Recall (from \cite{Susana_FSS04}) the definition of the \emph{ordinal powers} of a function
$G$ over a complete lattice $X$:

\[G \uparrow \alpha = \left\{
  \begin{array}{ll}
    \bigcup \{G \uparrow \alpha' \mid \alpha' < \alpha\}  
~~~~~~~ \mbox{ if } \alpha \mbox{ is a limit ordinal,} \\
G(G\uparrow (\alpha - 1))   ~~~~~~~~~~~~  \mbox{ if } \alpha
\mbox{ is a successor ordinal,}
  \end{array} \right.\]

\noindent and dually,

\[G \downarrow \alpha = \left\{
  \begin{array}{ll}
    \bigcap \{G \downarrow \alpha' \mid \alpha' < \alpha\} 
    ~~~~~~~  \mbox{ if }\alpha \mbox{ is a limit ordinal,} \\
    G(G\downarrow (\alpha - 1))  ~~~~~~~~~~~~  \mbox{ if } \alpha \mbox{
    is a successor ordinal,}
  \end{array} \right.\]

Since the first limit ordinal is 0, it follows that in particular,
$G \uparrow 0= \bot_X$ (the bottom element of the lattice $X$) and
$G \downarrow 0 = \top_X$ (the top element). From Kleene's fixed
point theorem we know that the least fixed-point of any continuous
operator is reached at the first infinite ordinal $\omega$. Hence
$lfp(T_P) = T_P \uparrow \omega$.


\begin{lem}
\label{lemma:prefixpoint}
  Let $P$ a fuzzy program. Then $M$ is a model of $P$ if and only if $M$ is a
  pre-fixpoint of $T_P$, that is $T_P(M) \sqsubseteq M$.
\end{lem}

\begin{pf}
  Let $M = \langle B_M,V_M \rangle$ and $T_P(M) = \langle B_{T_P},V_{T_P}
  \rangle$.

  We first prove the ``only if" ($\rightarrow$) direction. Let $A$ be an element of
  Herbrand Base, if $A \in B_{T_P}$, then by definition of $T_P$ there
  exists a ground instance of a fact of $P$, $A \gets v$, or a ground
  instance of a clause of $P$, $A \gets_F A_1,\ldots,A_n$ where
  $\{A_1,\ldots,A_n\} \subseteq B_M$ and $v =
  \mathcal{F}(V_M(A_1),\ldots,V_M(A_n))$. Since $M$ is a model of $P$,
  $A \in B_M$, and each $v \subseteq_{BI} V_M(A)$, then $V_{T_P}(A)
  \subseteq_{BI} V_M(A)$ and then $T_P(M) \sqsubseteq
  M$. $\square$. If $A \notin B_{T_P}$ then $V_{T_P}(A)= \mathit{default}(A)
  \subseteq_{BI} V_M(A)$.

  Analogously, for the ``if" ($\leftarrow$) direction, for each ground instance
  \\ $v = \mathcal{F}(V_M(A_1),\ldots,V_M(A_n))$, $A \in B_{T_P}$ and
  $v \subseteq_{BI} V_{T_P}(A)$, but as $T_P(M) \sqsubseteq M$, $B_{T_P}
  \subseteq B_M$ and $V_{T_P}(A) \subseteq_{BI} V_M(A)$. Then $A \in
  B_M$ and $v \subseteq_{BI} V_M(A)$ therefore $M$ is a model of
  $P$. $\square$
\end{pf}\\

Given this relationship, it is straightforward to prove that the
least model of a program $P$ is also the least fixed-point of
$T_P$.

\begin{thm}
\label{the:lmlfp}
  Let $P$ be a fuzzy program. Then $\mathit{lm}(P) = \mathit{lfp}(T_P)$.
\end{thm}

\begin{pf}
  $$
  \begin{array}{rcl}
 lm(P) & = & \bigcap \{ M \mid M \mbox{ is a model of } P \} \\
       & = & \bigcap \{ M \mid M \mbox{ is a pre-fixpoint of } P \} 
            \mbox{ from lemma \ref{lemma:prefixpoint}}\\
       & = & \mathit{lfp}(T_P) \ \ \ \mbox{ by the Knaster-Tarski} 
           \mbox{ Fixpoint Theorem
       \cite{tarski55:_lattic_theor}} \square
  \end{array}
$$
\end{pf}

\subsection{Operational Semantics}
\label{sec:opersem}

The improvement of Fuzzy Prolog is remarkable in its new procedural
semantics that is interpreted as a sequence of
transitions between different states of a system. We represent the
state of a \emph{transition system} in a computation as a tuple
$\langle A ,\sigma,S \rangle$ where $A$ is the goal, $\sigma$ is a
substitution representing the instantiation of variables needed to
get to this state from the initial one and $S$ is a constraint
that represents the truth value of the goal at this state.

When computation starts, $A$ is the initial goal, $\sigma =
\emptyset$ and $S$ is true (if there are neither previous
instantiations nor initial constraints). When we get to a state
where the first argument is empty then we have finished the
computation and the other two arguments represent the answer.

\begin{defn}[Transition]
A \emph{transition} in the \emph{transition system} is defined as:
\begin{enumerate}

\item  $\langle A \cup a,\sigma,S \rangle \to \langle
A\theta,\sigma \cdot \theta,S \land \mu_a =  v \rangle$

if $h \gets v$ is a fact of the program $P$, $\theta$ is the
mgu of $a$ and $h$, $\mu_a$ is the truth value for $a$ and
$\mathit{solvable}(S \land \mu_a = v )$.

\item $\langle A \cup a,\sigma,S \rangle \to \langle
(A \cup B)\theta,\sigma \cdot \theta,S \land c \rangle$

if $h \gets_F B$ is a rule of the program $P$, $\theta$ is
the mgu of $a$ and $h$, $c$ is the constraint that represents the
truth value obtained applying the union-aggregation $\mathcal{F}$
to the truth values of $B$, and $\mathit{solvable}(S \land c)$.

\item $\langle A \cup a,\sigma,S \rangle \to \langle A,\sigma,S \land
\mu_a = v\rangle$

if none of the above are applicable and $\mathit{solvable}(S \land \mu_a = v )$
where $\mu_a = \mathit{default}(a)$.
\end{enumerate}
\end{defn} 

\begin{defn}[Success set]
The success set $SS(P)$ collects the answers to simple goals
$p(\widehat{x})$. It is defined as follows: $~~~~~SS(P) = \langle B,V \rangle$ 

\noindent
where $B = \{ p(\widehat{x})\sigma | \langle
p(\widehat{x}),\emptyset,true \rangle \to^* \langle
\emptyset,\sigma,S \rangle \}$ is the set of elements of the
Herbrand Base that are instantiated and that have succeeded; and
$V(p(\widehat{x})) = \cup \{ v | \langle
p(\widehat{x}),\emptyset,true \rangle \to^* \langle
\emptyset,\sigma,S \rangle, \mbox{and } v \mbox{ is the solution
of }S \}$ is the set of truth values of the elements of $B$ that
is the union (got by backtracking) of truth values that are
obtained from the set of constraints provided by the program P
while query $p(\widehat{x})$ is computed.
\end{defn}









In order to prove the equivalence between operational semantic and
fixed-point semantic, it is useful to introduce a type of
canonical top-down evaluation strategy. In this strategy all
literals are reduced at each step in a derivation. For obvious
reasons, such a derivation is called \emph{breadth-first}.

\begin{defn}[Breadth-first transition]
 Given the following set of valid transitions:
$$ \begin{array}{lll} 
\langle \{\{A_1,\ldots,A_n\},\sigma,S
\rangle  \to  ~~~~~~~~~~ \langle
\{\{A_2,\ldots,A_n\}\cup B_1,\sigma \cdot \theta_1,S \land c_1
\rangle\\ 
\langle \{\{A_1,\ldots,A_n\},\sigma,S \rangle
\to  ~~~~~~~~~~ \langle \{\{A_1,A_3\ldots,A_n\}\cup
B_2,\sigma \cdot \theta_2,S \land c_2 \rangle \\
{~~~~~~~~~~~~~~~~~~~~~~~~~~~ \vdots }& \\ 
\langle
\{\{A_1,\ldots,A_n\},\sigma,S \rangle  \to  ~~~~~~~~~~
\langle \{\{A_1,\ldots,A_{n-1}\}\cup B_n,\sigma \cdot \theta_n,S
\land c_n \rangle \\
 \end{array}$$
 a \emph{breadth-first transition} is defined as

$\langle \{A_1,\ldots,A_n\},\sigma,S \rangle \to_{BF}$ 
\indent $ \langle B_1 \cup \ldots \cup B_n,\sigma \cdot \theta_1
\cdot \ldots \cdot \theta_n ,S \land c_1 \land \ldots \land c_n
\rangle $

\noindent in which all literals are reduced at one step.
\end{defn}




\begin{thm}
\label{the:succdertotp}
  Given a ordinal number $n$ and $T_P \uparrow n = \langle
   B_{T_{P_n}},V_{T_{P_n}} \rangle$.  There is a successful
   breadth-first derivation of lengh less or equal to $n + 1$ for a
   program $P$, $\langle \{A_1,\ldots,A_k\},\sigma,S_1 \rangle
   \to^{*}_{BF} \langle \emptyset,\theta,S_2 \rangle$ iff
   $A_i\theta \in B_{T_{P_n}}$ and $\mathit{solvable}(S \land \mu_{A_i} = v_i
   )$ and $v_i \subseteq_{BI} V_{T_{P_n}}(A_i\theta)$.
\end{thm}

\begin{pf}
  The proof is by induction on $n$. For the base case, all the
  literals are reduced using the first type of transitions or the last
  one, that is, for each literal $A_i$, it exits a fact $h_i \gets
  v_i$ such that $\theta_i$ is the mgu of $A_i$ and $h_i$, and
  $\mu_{A_i}$ is the truth variable for $A_i$, and $\mathit{solvable}(S_1 \land
  \mu_{A_i} = v_i )$ or $\mu_{A_i}=\mathit{default}(A_i)$. By definition of
  $T_P$, each $v_i \subseteq_{BI} V_{T_{P_1}}(A_i\theta)$ where
  $\langle B_{T_{P_1}},V_{T_{P_1}} \rangle = T_P \uparrow 1$.

  For the general case, consider the successful derivation, \\ $
  \langle \{A_1,\ldots,A_k\},\sigma_1,S_1 \rangle \to_{BF} \langle
  B,\sigma_2,S_2\rangle \to_{BF} \ldots \to_{BF} \langle
  \emptyset,\sigma_n,S_n \rangle $\\ the transition $\langle
  \{A_1,\ldots,A_k\},\sigma_1,S_1 \rangle \to_{BF} \langle
  B,\sigma_2,S_2\rangle$
  
  When a literal $A_i$ is reduced using a fact or there is not rule
  for $A_i$, the result is the same as in the base case. Otherwise
  there is a clause $h_i \gets_F B_{1_i},\ldots,B_{m_i}$ in $P$ such
  that $\theta_i$ is the mgu of $A_i$ and $h_i \in B\sigma_2$ and
  $B_{j_i}\theta_i \in B\sigma_2$, by the induction hypothesis
  $B\sigma_2 \subseteq B_{T_{P_{n-1}}}$ and $\mathit{solvable}(S_2 \land
  \mu_{B_{j_i}} = v_{j_i} )$ and $v_{j_i} \subseteq_{BI}
  V_{T_{P_{n-1}}}(B_{j_i}\sigma_2)$ then $B_{j_i}\theta_i \subseteq
  B_{T_{P_{n-1}}}$ and by definition of $T_P$, $A_i\theta_i \in
  B_{T_{P_n}}$ and $\mathit{solvable}(S_1 \land \mu_{A_i} = v_i )$ and $v_i =
  \subseteq_{BI} V_{T_{P_n}}(A_i\sigma_1)$. $\square$
 \end{pf}

\begin{thm}
\label{the:ssplfp}
For a program $P$ there is a successful
derivation $$\langle p(\widehat{x}),\emptyset,true \rangle
\to^* \langle \emptyset,\sigma,S \rangle$$ iff
$p(\widehat{x})\sigma \in B$ and $v \mbox{ is the solution of }S$
and $v \subseteq_{BI} V(p(\widehat{x})\sigma)$ where $lfp(T_P)=
\langle B,V\rangle$
\end{thm}

\begin{pf} It follows from the fact that $lfp(T_P) = T_P \uparrow
  \omega$ and from the Theorem  \ref{the:succdertotp}. $\square$
\end{pf}

\begin{thm} For a fuzzy program $P$ the three semantics
are equivalent, i.e.
\[SS(P) = lfp(TP) = lm(P)\]
\end{thm}

\begin{pf}
  the first equivalence follows from Theorem \ref{the:ssplfp} and the
  second from Theorem \ref{the:lmlfp}. $\square$
\end{pf}

\section{Implementation and Syntax}
\label{sec:clpr}

\subsection{CLP(${\cal R}$)}

Constraint Logic Programming \cite{jaff87-Cons} began as a natural
merging of two declarative paradigms: constraint solving and logic
programming. This combination helps make CLP programs both
expressive and flexible, and in some cases, more efficient than
other kinds of logic programs.  CLP(${\cal R}$)
\cite{JaffarMichaylovStuckeyYap92TOPLAS} has linear arithmetic
constraints and computes over the real numbers.

Fuzzy Prolog was implemented in \cite{Susana_FSS04} as a syntactic
extension of a CLP(${\cal R}$) system. CLP(${\cal R}$) was
incorporated as a library in the Ciao Prolog system\footnote{The Ciao
system \cite{ciao-modules-cl2000} including our Fuzzy Prolog
implementation can be downloaded from
http://www.clip.dia.fi.upm.es/Software/Ciao.}.



The \emph{fuzzy} library (or \emph{package} in the Ciao Prolog
terminology) which implements the interpreter of our Fuzzy Prolog
language has been modified to handle default reasoning.

\subsection{Syntax}

Let us recall, from \cite{Susana_FSS04}, the syntax of Fuzzy Prolog. Each
Fuzzy Prolog \emph{clause} has an additional argument in the head which
represents its truth value in terms of the truth values of the subgoals of the
body of the clause. A \emph{fact} $A \gets v$ is represented by a Fuzzy Prolog
fact that describes the range of values of $v$ with a union of intervals
(which can be only an interval or even a real number in particular cases). The
following examples illustrate the concrete syntax of programs:

\noindent
\begin{tabular}{ll}
$\mathit{youth}(45) \gets ~[0.2,0.5] ~\bigcup~ [0.8,1]~~~~~~~~~~~~~$ &    {\tt youth(45):{\tiny$^\sim$} [0.2,0.5]v[0.8,1]}\\

$\mathit{tall}(john) \gets 0.7$ & {\tt tall(john):{\tiny$^\sim$} 0.7.}
\\

$\mathit{swift}(john) \gets ~ [0.6,0.8]$  & {\tt swift(john):{\tiny$^\sim$} [0.6,0.8] }\\

$\mathit{good\_player}(X) \gets_{min}~ \mathit{tall}(X),$   & {\tt
good\_player(X):{\tiny$^\sim$}min  tall(X),} \\

$~~~~~~~~~~~~~~~~~~~~~~~~~~~~~~ \mathit{swift}(X)$ \  & $~$ {\tt ~~~~~~~~~~~~~~~~~~ swift(X)}\\

\end{tabular}

These clauses are expanded at compilation time to constrained
clauses that are managed by CLP(${\cal R}$) at run-time.
Predicates $\mathit{.=.} /2$, $\mathit{.<.} /2$, $\mathit{.<=.} /2$, $\mathit{.>.} /2$ and $\mathit{.>=.} /2$
are the Ciao CLP(${\cal R}$) operators for representing constraint
inequalities, we will use them in the code of predicates definitions (while we
will use the common operators $=$, $<$, $\leq$, $>$, $\geq$ for theoretical definitions).  For example the first fuzzy fact is expanded to
these Prolog clauses with constraints

\begin{verbatim}
youth(45,V):-   V .>=. 0.2, V .<=. 0.5.
youth(45,V):-   V .>=. 0.8, V .<.  1.
\end{verbatim}

And the fuzzy clause

\noindent
\begin{quotation} \tt
good\_player(X) {\tt:{\tiny$^\sim$}} min tall(X), swift(X).
\end{quotation}

is expanded to

\begin{verbatim}
good_player(X,Vp) :- tall(X,Vq), swift(X,Vr), minim([Vq,Vr],Vp), 
                     Vp .>=. 0, Vp .=<. 1.
\end{verbatim}

The predicate \texttt{minim/2} is included as run-time code by the
library. Its function is adding constraints to the truth value
variables in order to implement the T-norm \emph{min}.\\

\begin{minipage}{2in}
\begin{verbatim}
minim([],_).
minim([X],X).
minim([X,Y|Rest],Min):- min(X,Y,M), minim([M|Rest],Min).

min(X,Y,Z):- X .=<. Y , Z .=. X.
min(X,Y,Z):- X .>. Y, Z .=. Y .
\end{verbatim}
\end{minipage}\\

We have implemented several \emph{aggregation operators} as
\texttt{prod}, \texttt{max}, \texttt{luka} (Lukasievicz operator), etc. and in a similar
way any other operator can be added to the system without any
effort. The system is extensible by the user simply adding the
code for new \emph{aggregation operators} to the library.



\section{Combining Crisp and Fuzzy Logic}
\label{sec:comb}

\subsection{Example: Teenager Student}

In order to use definitions of fuzzy predicates that include crisp subgoals we
must define properly their semantics with respect to the Prolog Close World
Assumption (CWA) \cite{Clark}. We are going to present a motivating
example from \cite{Susana_FSS04}.

Fuzzy clauses usually use crisp predicate calls as requirements that
data have to satisfy to verify the definition in a level superior to
0, i.e. crisp predicates are ussually tests that data should satisfy in the
body of fuzzy clauses. For example, if we can say that a teenager student is a student
whose age is about 15 then we can define the fuzzy predicate
$teenager\_student/2$ in Fuzzy Prolog as 
\begin{verbatim}
teenager_student(X,V):~ student(X), age_about_15(X,V2).
\end{verbatim}


Note that we can face the risk of unsoundness unless the semantics of
crisp and fuzzy predicates is properly defined. CWA means that all
non-explicit information is false.  E.g., if we have
the predicate definition of $student/1$ as
\begin{verbatim}
student(john).  
student(peter).
\end{verbatim}
then we have that the goal $student(X)$ succeeds with $X=john$ or
with $X=peter$ but fails with any other value different from
these; i.e:
\begin{verbatim}
?- student(john).
yes

?- student(nick).
no
\end{verbatim}

which means that $john$ is a student and $nick$ is not. This is the
semantics of Prolog and it is the one we are going to adopt for crisp
predicates because we want our system to be compatible with
conventional Prolog reasoning. But what about fuzzy predicates? According to
human reasoning we should assume OWA (non explicit information in unknown).
Consider the following definition of $age\_about\_15/2$  
\begin{verbatim}
age_about_15(john,1):~ . 
age_about_15(susan,0.7):~ .
age_about_15(nick,0):~ .
\end{verbatim}
\noindent
The goal $age\_about\_15(X,V)$ succeeds with $X=john$ and $V=1$
or with $X=susan$ and $V=0.7$. 
Therefore we do not know if the age of $peter$ is
about 15 or not; and we know that $nick$'s age is definitely not about 15.

Our way to introduce crisp subgoals into the body of fuzzy clauses is
translating the crisp predicate into the respective fuzzy predicate. For our
example we obtain the following Prolog definition.
\begin{verbatim}
teenager_student(X,V):~ f_student(X,V1), age_about_15(X,V2).
\end{verbatim}

\noindent
Where the
default truth value of a crisp predicate is $0$.

\begin{verbatim}
f_student(X,1):- student(X).
:-default(f_student/2,0).
\end{verbatim} 
Nevertheless, we consider for $age\_about\_15/2$ and $teenager\_student/2$ that
the default value is unknown (the whole interval $[0,1]$).

\begin{verbatim}
:-default(age_about_15/2,[0,1]).
:-default(teenager_student/2,[0,1]).
\end{verbatim} 

Observe the following consults:

\begin{verbatim}
?- age_about_15(john,X).
X = 1

?- age_about_15(nick,X).
X = 0 

?- age_about_15(peter,X).
X .>=. 0, X .<=. 1  
\end{verbatim}

This means $john$'s age is about 15, $nick$'s age is not about 15 and
we have no data about $peter$'s age.

We expect the same behavior with the fuzzy predicate
$teenager\_student/2$, ie:
\begin{verbatim}
?- teenager_student(john,V).
V .=. 1

?- teenager_student(susan,V).
V .=. 0

?- teenager_student(peter,V).
V .>=. 0, V .<=. 1  
\end{verbatim}

as $john$ is a ``teenager student'' (he is a student and his age is
about 15), $susan$ is not a ``teenager student'' (she is not a
student) and we do not know the value of maturity of $peter$ as
student because although he is a student, we do not know if his age is
about 15. 





\subsection{Example: Timetable Compatibility}

\begin{figure*}
        \begin{center}
                \includegraphics[totalheight=5.2 cm]{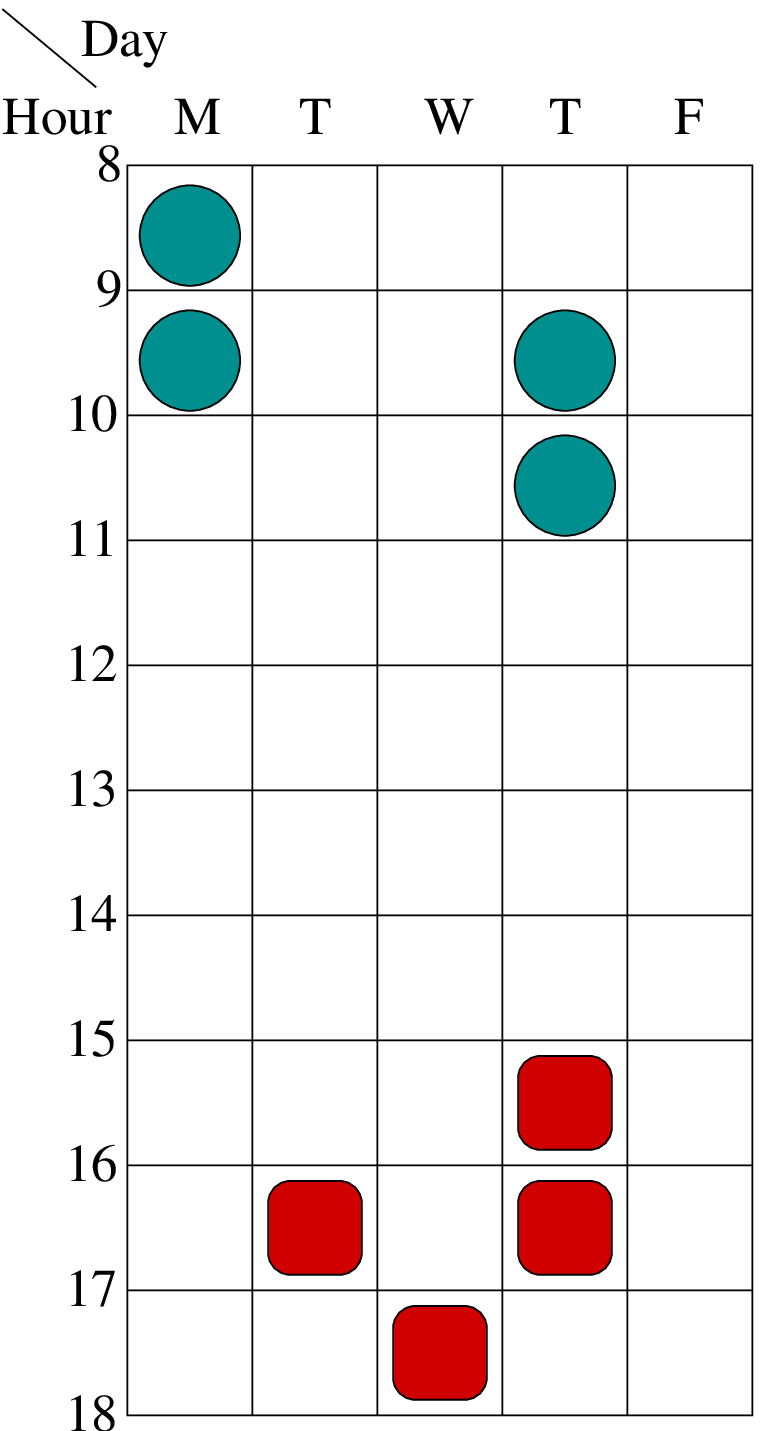}
                \includegraphics[totalheight=5.2 cm]{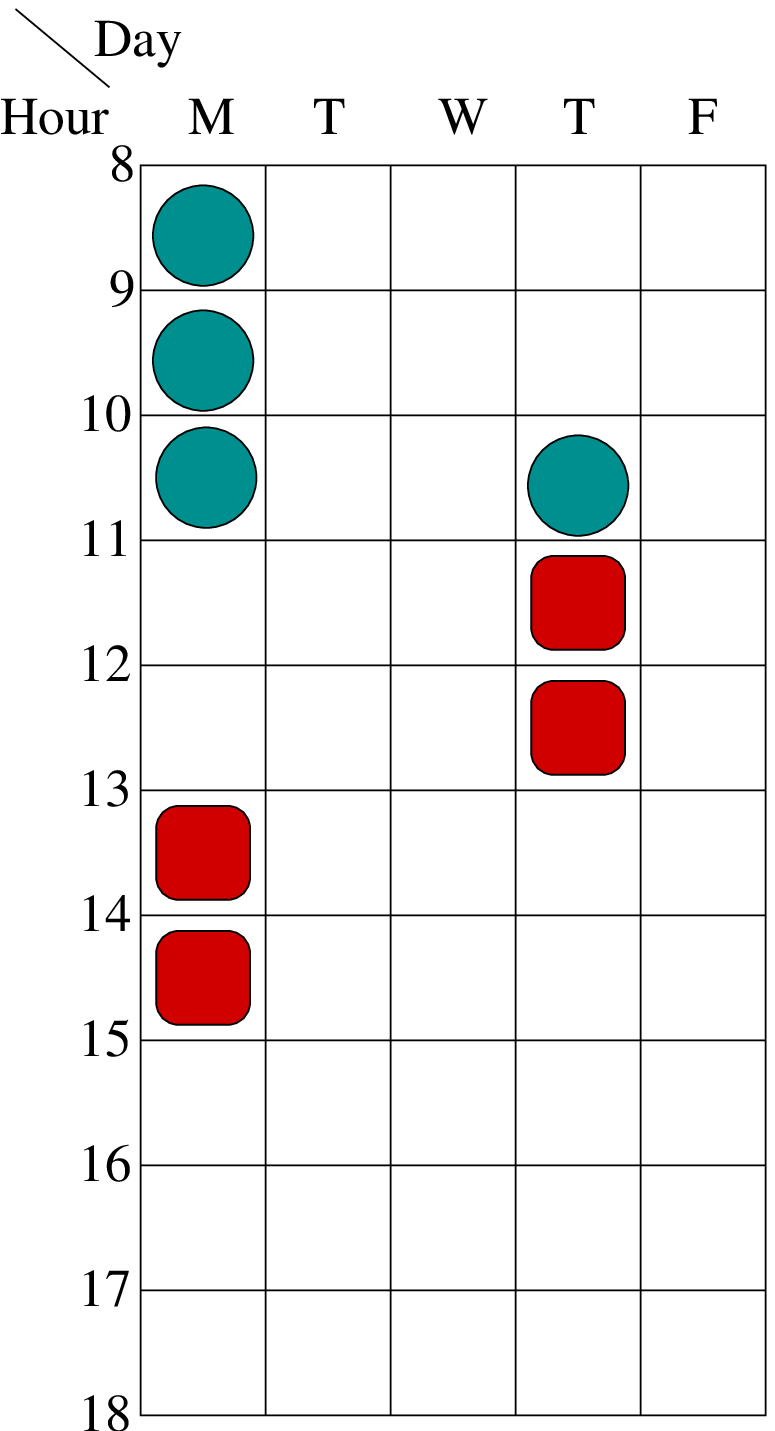}
                \includegraphics[totalheight=5.2 cm]{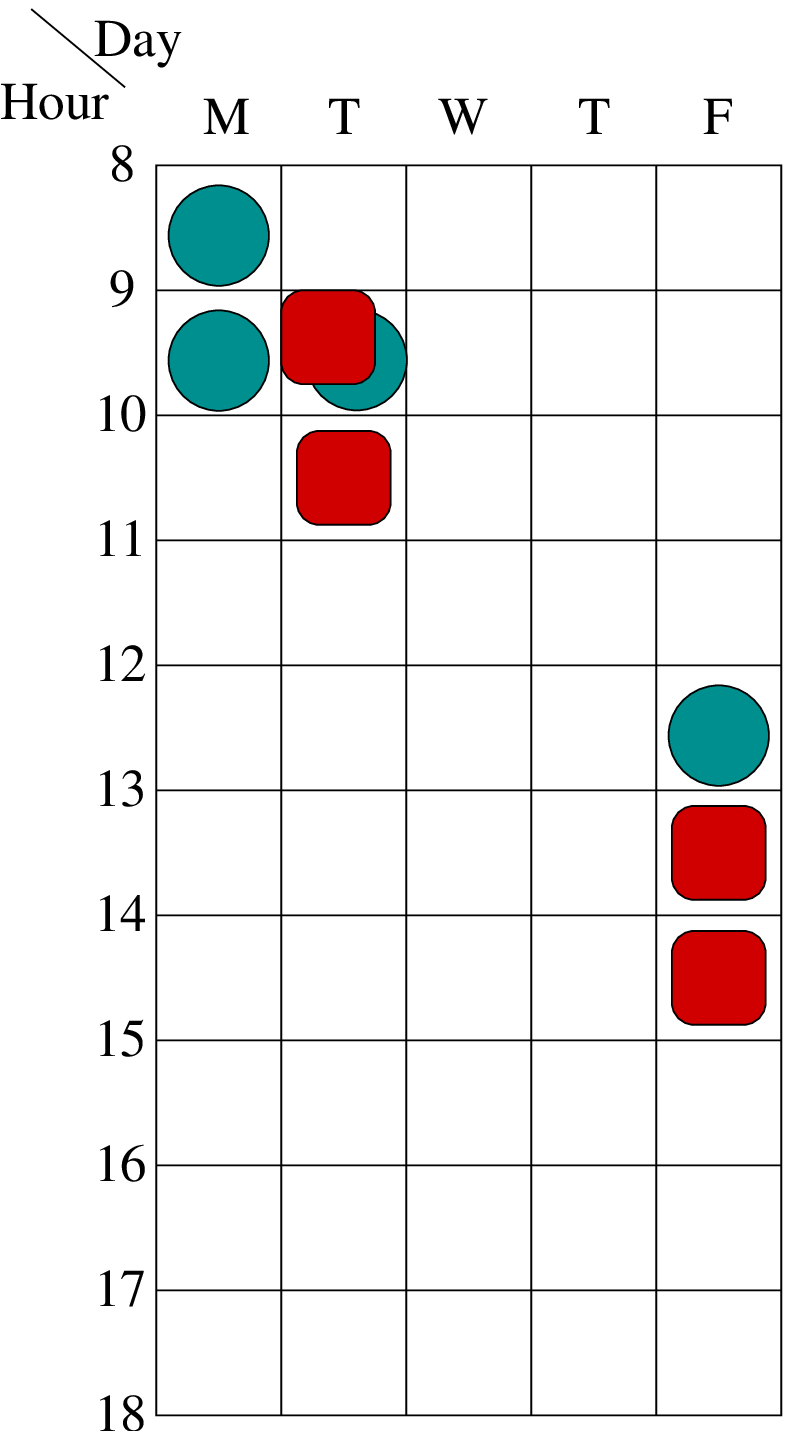}
                \includegraphics[totalheight=5.2 cm]{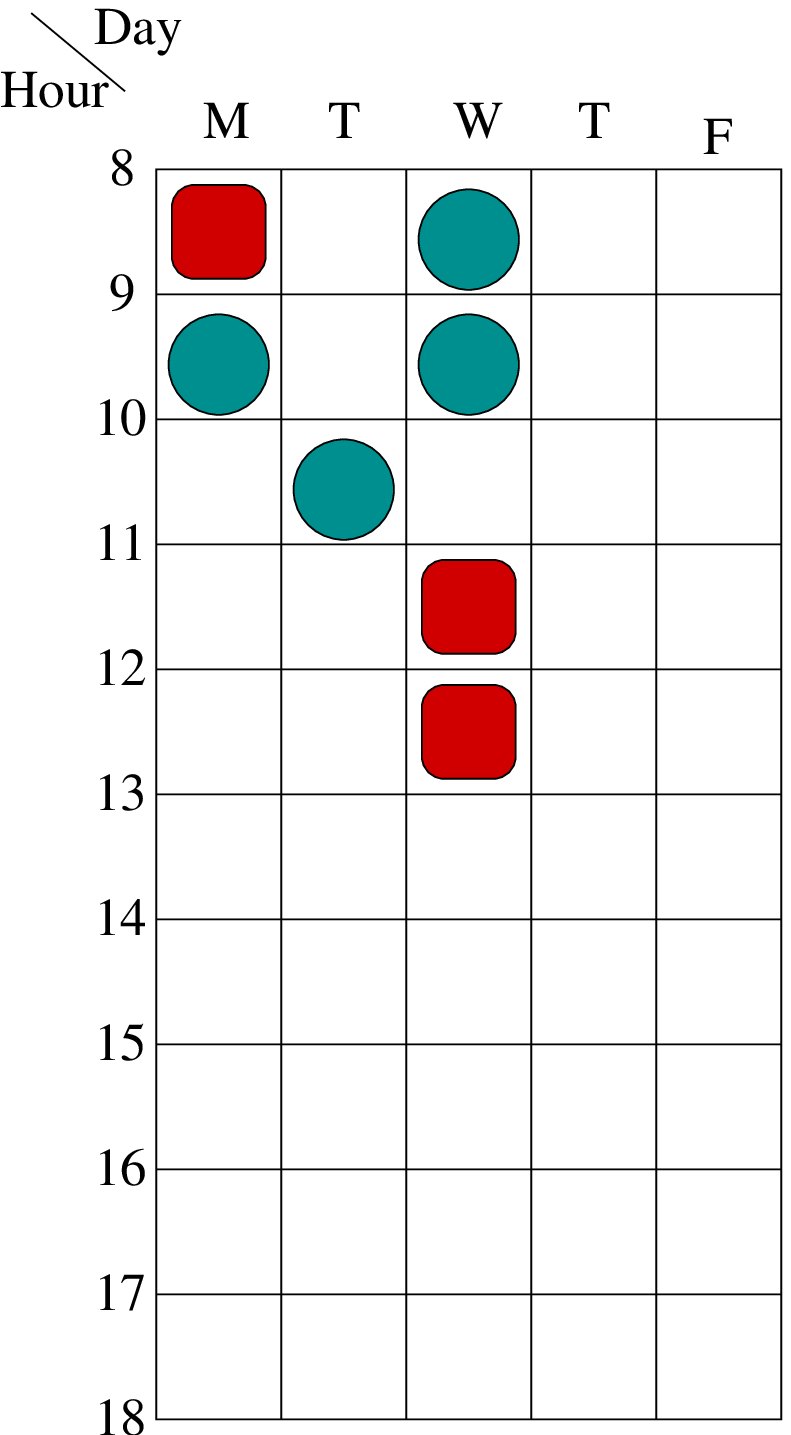}
        \end{center}
        \caption{Timetable 1, 2, 3 and 4}
        \label{fig:timetables}
\end{figure*}

Another real example could be the problem of compatibility of a couple
of shifts in a work place. For example teachers that work in different
class timetables, telephone operators, etc. Imagine a company where
the work is divided in shifts of 4 hours per week. Many workers have to
combine a couple of shifts in the same week and a predicate
$compatible/2$ is necessary to check if two shifts are compatible or to
obtain which couples of shifts are compatible. Two shifts are compatible
when both are correct (working days from Monday to Friday, hours
between 8 a.m. and 18 p.m. and there are no repetitions of the same
hour in a shift) and in addition when the shifts are disjoint.

\begin{verbatim}
compatible(T1,T2):- correct_shift(T1), correct_shift(T2), 
                    disjoint(T1,T2).
\end{verbatim}

But there are so many compatible combinations of shifts that it would
be useful to define the concept of compatibility in a fuzzy way
instead of in the crisp way it is defined above. It would express that
two shifts could be incompatible if one of them is not correct or if
they are not disjoint but when they are compatible, they can be more
or less compatible. They can have a level of compatibility. Two shifts
will be more compatible if the working hours are concentrated (the
employee has to go to work few days during the week). Also, two shifts
will be more compatible if there are few free hours between the busy
hours of the working days of the timetable.

\begin{figure*}
        \begin{center}
                \includegraphics[totalheight=3.2cm]{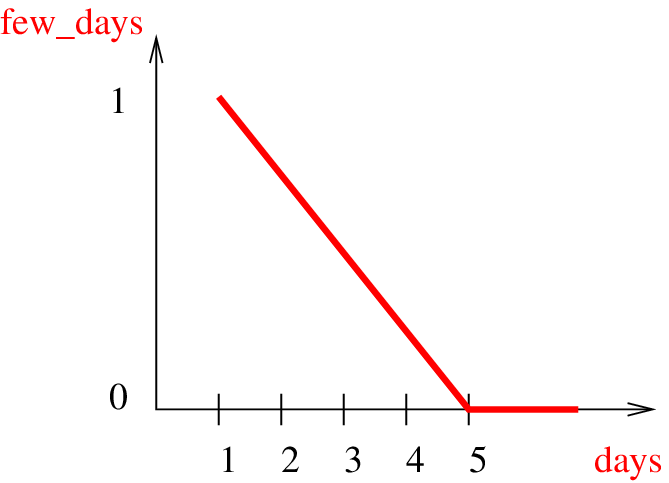}
                \includegraphics[totalheight=3.2cm]{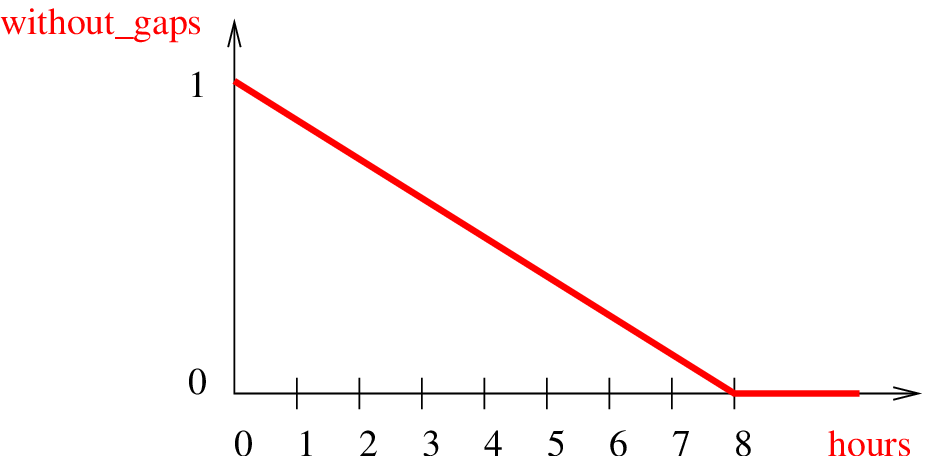}
        \end{center}
        \caption{Fuzzy predicates few\_days/2 and without\_gaps/2}
        \label{fig:predicates}
\end{figure*}

Therefore, we are handling crisp concepts ($correct\_shift/1$,
$disjoint/2$) besides fuzzy concepts ($without\_gaps/2$,
$few\_days/2$). Their definitions, represented in figure
\ref{fig:predicates}, are expressed in our language in this simple
way (using the operator ``$\mathit{:\#}$'' for function definitions and the
reserved word ``$\mathit{fuzzy\_predicate}$''):

\begin{verbatim}
few_days :# 
   fuzzy_predicate([(0,1),(1,0.8),(2,0.6),(3,0.4),(4,0.2),(5,0)]).

without_gaps :# 
   fuzzy_predicate([(0,1),(1,0.8),(5,0.3),(7,0.1),(8,0)]).
\end{verbatim}

A simple implementation in Fuzzy Prolog combining both types of predicates
could be:
\begin{verbatim}
compatible(T1,T2,V):~ min
        f_correct_shift(T1,V1),
        f_correct_shift(T2,V2),
        f_disjoint(T1,T2,V3),
        f_append(T1,T2,T,V4),
        f_number_of_days(T,D,V5),
        few_days(D,V6),
        f_number_of_free_hours(T,H,V7),
        without_gaps(H,V8).
\end{verbatim}

Here $append/3$ gives the total weekly timetable of 8 hours from joining two
shifts, $number\_of\_days/3$ obtains the total number of working days of a
weekly timetable and $number\_of\_free\_hours/2$ returns the number of free
one-hour gaps that the weekly timetable has during the working days. The
$\verb#f_#\mathit{predicates}$ are the corresponding fuzzified crisp predicates. The
aggregation operator $min$ will aggregate the value of $V$ from $V6$ and
$V8$ checking that $V1$, $V2$, $V3$, $V4$, $V5$ and $V7$ are equal to $1$,
otherwise it fails.  Observe the timetables in figure \ref{fig:timetables}. We
can obtain the compatibility between the couple of shifts, T1 and T2,
represented in each timetable asking the subgoal $compatible(T1, T2, V)$. The
result is $V=0.2$ for the timetable 1, $V=0.6$ for the timetable 2, and
$V=0$ for the timetable 3 (because the shifts are incompatible).

Regarding
compatibility of shifts in a weekly timetable, we are going to ask
some questions about the shifts T1 and T2 of timetable 4 of figure
\ref{fig:timetables}. One hour of T2 is not fixed yet.

We can note: the days of the week as $mo$, $tu$, $we$, $th$ and $fr$;
the slice of time of one hour as the time of its beginning from $8$
a.m. till $17$ p.m.; one hour of the week timetable as a pair of day
and hour and one shift as a list of 4 hours of the week.



If we want to know how to complete the shift T2 given a level of
compatibility higher than 70 \%, we obtain the slice from 10 to 11
p.m. at Wednesday or Monday morning.  
\begin{verbatim}
?- compatible([(mo,9), (tu,10), (we,8), (we,9)],
              [(mo,8), (we,11), (we,12), (D,H)], V), 
   V .>. 0.7 .

V = 0.9,  D = we, H = 10 ? ;
V = 0.75, D = mo, H = 10 ? ;
no
\end{verbatim}


\section{Conclusions and Future work}
\label{sec:concl}

Extending the expressivity of programming systems is very important for
knowledge representation. We have chosen a practical and extended language for knowledge
representation: Prolog.

Fuzzy Prolog presented in \cite{Susana_FSS04} is implemented over Prolog
instead of implementing a new resolution system. This gives it a good
potential for efficiency, more simplicity and flexibility. For example
\emph{aggregation operators} can be added with almost no effort. This
extension to Prolog is realized by interpreting fuzzy reasoning as a set of
constraints \cite{Zad1}, and after that, translating fuzzy predicates into
CLP(${\cal R}$) clauses. The rest of the computation is resolved by the
compiler.

In this paper we propose to enrich Prolog with more expressivity by adding
default reasoning and therefore the possibility of handling incomplete
information that is one of the most worrying characteristics of data
(i.e. all information that we need usually is not available but only one part
of the information is available) and anyway searches, calculations,
etc. should be done just with the information that we had.

We have developed a complete and sound semantics for handling incomplete fuzzy
information and we have also provided a real implementation based in our
former Fuzzy Prolog approach.

We have managed to combine crisp information (CWA) and fuzzy information (OWA
or default) in the same program. This is a great advantage because it lets us
model many problems using fuzzy programs. So we have extended the expressivity
of the language and the possibility of applying it to solve real problems in
which the information can be defined, fuzzy or incomplete.

Presently we are working in several related issues:
\begin{itemize}
        \item Obtaining constructive answers to negative goals.
        \item Constructing the syntax to work with discrete fuzzy sets
        and its applications (recently published in \cite{SusanaPADL2005}).
        \item Implementing a representation model using unions instead of
        using backtracking.
        \item Introducing domains of fuzzy sets using types. This seems to be
        an easy task considering that we are using a modern Prolog \cite{CIAO}
        where types are available.
        \item Implementing the expansion over other systems. We are studing
        now the advantages of an implementation in XSB system where tabling is
        used.
	\item Using our approach for the engine of robots in a RoboCup league
	in a joint project between our universities.
\end{itemize}



 
\bibliographystyle{plain}  
\bibliography{bib_default}

\end{document}